# Heterogeneous local plastic deformation of interstitial free steel revealed using *in-situ* tensile testing and high angular resolution electron backscatter diffraction

J L R Hickey[1], S Rouland[2] & T B Britton[1*]

[1]Department of Materials, Imperial College London, Prince Consort Road, London, SW7 2AZ, UK.

[2]Electricité de France, R&D Division, Material and Mechanics of Components Department, Les Renardières, 77818 Moret-sur Loing Cedex, France.

*Corresponding Author: T.B. Britton, b.britton@imperial.ac.uk / Department of Materials, Imperial College London, Prince Consort Road, London, SW7 2AZ, UK.

ORCiDs: Jim Hickey - 0000-0002-7845-020X; Solene Rouland - 0000-0002-0084-5850; Ben Britton - 0000-0002-7845-020X

## Abstract

Metals are important structural materials for transport and the built environment. Low carbon steels can fail through strain localisation due to the role of interstitial solute atoms (such as carbon and nitrogen) interacting with mobile dislocations, and this gives rise to the Portevin–Le Chatelier effect and the formation of Lüders bands. In this work, we use *in-situ* tensile testing and observation with High Angular Resolution Electron Back Scatter Diffraction (HR-EBSD) to explore deformation patterning in two Interstitial Free (IF) steel samples. One of these steels was heat treated to trigger strain localisation and Lüders bands whilst the other was heat treated to homogenise plastic strain and limit flow localisation. Our work reveals that flow localisation at the macroscopic scale is closely correlated with differences in the storage of dislocations at the microscale through analysis of the HR-EBSD derived fields of stored Geometrically Necessary Dislocations (GNDs). Homogeneous plastic flow correlates with more 'Face Centred Cubic (FCC)-like' deformation patterning in these Body Centred Cubic (BCC) materials, where work hardening is correlated with the association of dislocation networks which interact with triple junctions and grain boundaries, and our *in-situ* tests enable us to see how these fields develop. Our findings enable the modification of steel processing routes, and micromechanical models, based upon information obtained from these *in-situ* tests.

# 1 Introduction

Global annual production of steel is approximately 1600 million metric tons [1] and this material underpins modern society as a critical building material in the built environment and the underpinning structural material in cars and ships. In the automotive industry, the skin of most cars is made of sheet Interstitial Free (IF) steel as the presence of mobile interstitials presents a significant risk of flow localisation through the generation of Lüders bands. Typically, Materials Scientists and Engineers consider that these bands are formed due to the impact of mobile interstitial solute atoms such as carbon and nitrogen which pins dislocation sources and mobile dislocations [2]. Once these dislocations are freed from the carbon and nitrogen interstitial traps, they are highly mobile and the local region is softer than the remainder of the material. This amplifies the relative dislocation activity in this local region causing strain localisation which macroscopically is observed as a strain band. These bands make sheet forming extremely difficult and can create significant scrap in the automobile production process. Understanding the flow localisation process motivates this present work.

Flow localisation at the atomic scale has been considered in simulations [3–8] and experiments [9–11] but it is difficult to link these processes together to understand how flow localisation, load drops and Lüders bands relate from atomic scale mechanisms up towards polycrystals. There must be a link between the motion and storage of dislocations (at the nm to µm length scale) and their interactions with microstructure, namely grain boundaries and triple junctions that determines how these materials deform. We have opted to explore this interlink through employment of *in-situ* tensile testing and repeated measurements, under load, of the stored dislocation content using High Angular Resolution Electron Backscatter Diffraction (HR-EBSD).

HR-EBSD[12,13] has been used to quantify the micron-scale deformation patterns of metals [14–20], semiconductor [21–24] and geological materials [25,26]. One notable advantage of this technique is its ability to combine a spatial resolution of better than 0.1 µm with strain and rotation gradients resolutions of $10^{-4}$ and $10^{-4}$ rads respectively [27]. This allows resolution of features with high spatial, strain and angular resolution over a large field of view and in a reasonable time frame, thanks to other advancements such as fast and high resolution EBSD cameras.

Cross-correlation of Electron Backscatter Diffraction Patterns (EBSPs) enables measurements in changes of the displacement gradient between points in a map. These fields can be spatially differentiated and used to calculate maps of Geometrically Necessary Dislocation (GND) density [16,18–20,26,28–30], and these GND density fields can also be calculated from Hough-based EBSD [31]. Studying the development of a GND density field offers insight into how the microstructure is deforming as a function of applied strain, texture, grain morphology and networked microstructure. For example, Jiang et al. [19] studied the effect of step size and EBSD camera binning on the resultant GND density field characteristics. They found that binning the detector to 4 x 4 (250 x 250 pixels in their case) and a step-size of 0.5 µm resulted in an acceptable angular resolution from the cross-correlation of the EBSPs of copper to systematically evaluate the GND density. In practice, GND density field calculations will always be a compromise between experiment parameters such as the approximate dislocation cell size [32], map step-size, required map size and instrument time, as changing the step-size alters the size of the measured Burgers circuit resulting in a decrease in the measured GND density [19].

Advancement in *in-situ* mechanical testing combined with HR-EBSD has enabled measurement of the elastic strain gradient in micron-scale silicon tensile specimens [33], as well as the GND density in a loaded tungsten cantilever beam [34]. *In-situ* deformation has notable advantages over *ex-situ* techniques such as: (1) the requirement that the sample does not need to be moved or fully unloaded to take an EBSD scan; (2) ease of accessing the same Area of Interest (AOI) leading to an efficient tracking capability of the same area with increasing strain.

IF steel [35–37] is a low strength, high ductility steel that consists of a ferrite matrix with titanium or niobium carbides in the microstructure. When recrystallizing from a rolled plate at approximately 750ºC – 900ºC, the rolled {111}||ND and {110}||RD/TD (commonly referred to as the γ and α fibres) are retained [38]. The processing and heat-treatments can redistribute solute elements such as carbon and nitrogen in IF steel in order to adjust the initial mobile dislocation content [35,39–41]. This can result in sufficient dislocation immobility, which usually manifests as a discontinuous yield when the steel is plastically deformed [2].

The objectives of this work are to therefore: (1) demonstrate the value of combined HR-EBSD with *in-situ* tensile testing to measure GND density fields on the same AOI; (2) understand the deformation response of two samples of IF steel manufactured from the same rolled plate and subjected to two different heat-treatments; (3) correlate these *in-situ* observations with the macroscopic *ex-situ* deformation response.

## 2 Method

Table 1 shows the composition of the as-received IF steel used in this study.

Table 1 – Composition of IF steel used in this study in wt%. The carbon and sulphur contents were determined using a combustion infrared detection technique. The nitrogen content was determined using the inert gas fusion technique. All other elements were determined using the inductively coupled plasma optical emission spectroscopy technique.

| Fe  | C     | Mn   | P    | S     | N     | Si   | Cu   | Ni    | Cr   | Ti    |
|-----|-------|------|------|-------|-------|------|------|-------|------|-------|
| Bal | 0.007 | 0.20 | 0.01 | 0.001 | 0.002 | 0.03 | 0.03 | <0.01 | 0.02 | <0.01 |

Four samples of as-received IF steel were split into two groups. One group was heat-treated for 4 hours at 700°C followed by 1 hour at 900°C and air cooled. The second group was heat-treated using the same steps except a 24 hour, 900°C soak time as shown in Figure 1(a). The longer soak time would result in a change in the distribution of the solute elements in each microstructure[35].

A Gatan MTest 2000E tensile tester was used to deform the samples. The contrast for DIC was generated on the gauges of the samples by grinding them to a 4000 grit finish, spraying them with white paint and depositing photocopying carbon toner [42]. An in-house cross-correlation-based DIC code was used that resulted in a strain component field spatial resolution of approximately 30 µm.

Separate samples from both heat-treatments were subsequently deformed *in-situ* using the tensile frame but now within a FEI Quanta 650g FEG-SEM. The samples were ground to a 4000-grit finish and Ar ion-beam polished using a Gatan PECS II. The final step in the ion beam polishing protocol was a beam angle of 1º relative to the sample surface at 1 kV. Micro indents were used to define an AOI of approximately 500 µm x 300 µm in the centre of the tensile specimen gauges. As-received texture and grain morphology maps were taken using Hough-based EBSD with a step size of 1 µm, presented in Figure 1(b)-(d). EBSD measurements were captured with a beam tension of 20 kV and probe current of ~10 nA. Due to time constraints, a smaller AOI of 200 µm x 150 µm was then defined within the initial region and a second map was taken of this region at a step size of 0.35 µm, an EBSD detector binning of 4x4 (i.e. EBSP resolution of 400 x 300 pixels) and an EBSD camera exposure time of 45 ms to capture the HR-EBSD data. The step size was calculated from an estimation of the dislocation cell size of 0.7 µm in low carbon steels at an average dislocation density of $10^{14}$ m$^{-2}$ using the method presented in [43].

The texture and grain morphology as a result of both heat-treatments are presented as Crystal Orientation Maps (COM), coloured with Inverse Pole Figure (IPF) colours with respect to the tensile axis (x) and {111} and {110} pole figures in Figure 1.(b)-(d) and were plotted using MTex 5.0 [44].

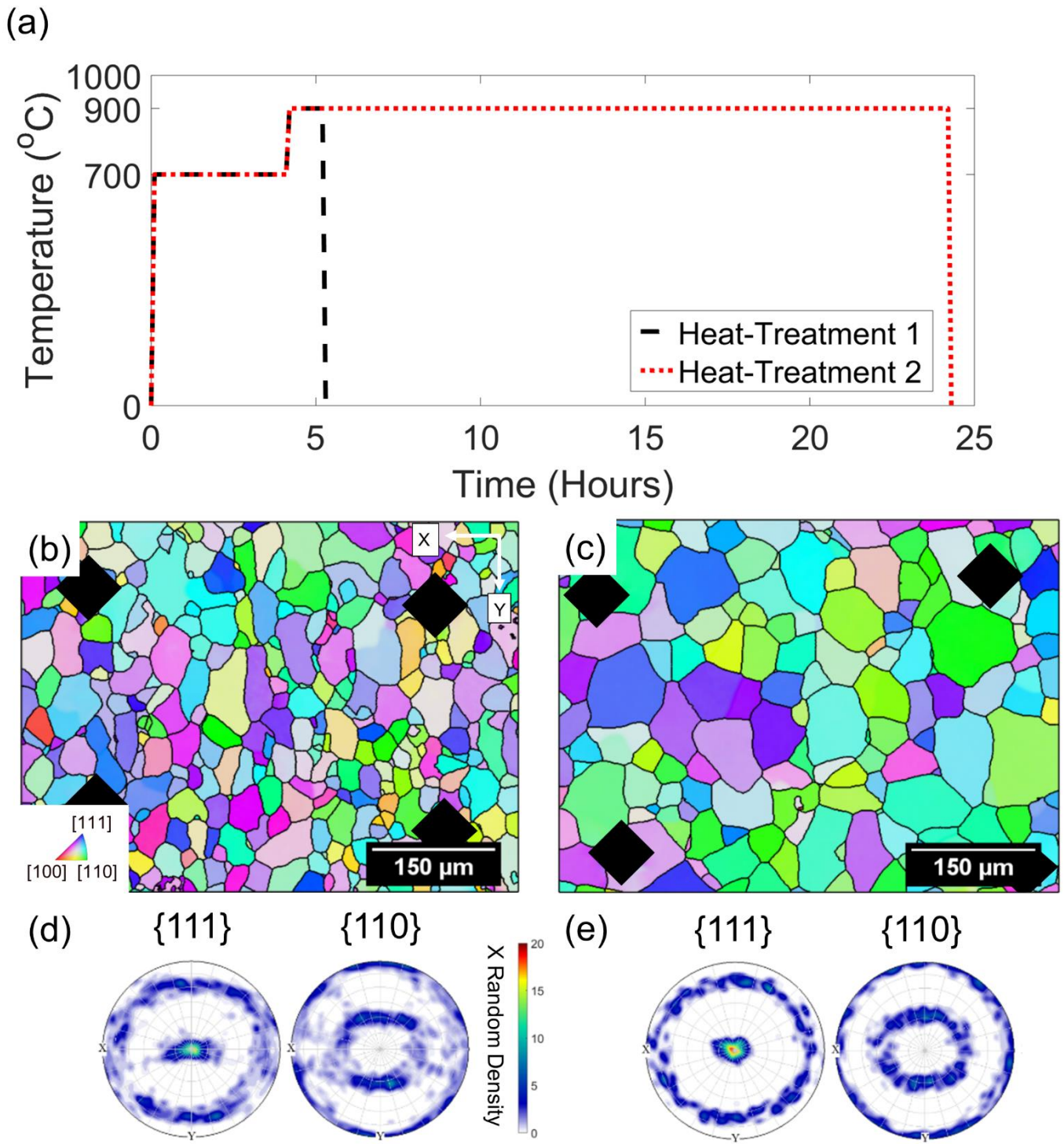

*Figure 1 – (a) Plot of temperature Vs time for both heat-treatments used in this study and (b) & (c) crystal orientation maps w.r.t. the x (tensile) axis of the defined AOI on the gauges of the samples subjected to heat-treatment 1 & 2 respectively prior to any deformation; (d) & (e) corresponding pole figures comparing the initial microstructures & deformation response of the samples subjected to heat-treatments 1 & 2. [For interpretation of the colour figures, please see the online version]*

The samples were then deformed to strains of 0.002, 0.02 and 0.04. At each strain, the sample was unloaded to 75% of the applied stress value to arrest significant creep effects and a further HR-EBSD scan was taken using the same parameters as the 0 strain case. The GND density fields were then calculated using an in-house EBSP cross-correlation code using the method described in [45] that utilises the Nye tensor [46], an L1 minimisation scheme [29] and the {110}<111> slip system in ferrite.

After deformation, EBSD mapping was performed (to check the AOI) and an Energy Dispersive X-Ray Spectroscopy (EDS) map (400 μm x 300 μm) was simultaneously captured for qualitative EDS-based distinction of the Ti-based precipitate distributions. EDS was performed at 20 kV using a Bruker XFlash 6 detector and the Ti peak was extracted to produce the maps in Figure 2.

## 3 Results & Discussion

All experiments were performed on the same batch of cold-rolled IF steel subjected to two different heat-treatments. The aim of the first heat-treatment (heat-treatment 1) was to recrystallize the sample and produce stress-drops around the yield-point as a result of Cottrell atmosphere-based[2] pinning of dislocations. The aim of the second heat-treatment (heat-treatment 2) was to recrystallize the sample and produce no stress-drops. The heat-treatments were also designed to have similar texture and grain morphologies. As shown in Figure 1(a), both heat-treatments differed significantly only by the longer 900°C soak time for heat-treatment 2. The arithmetic means of the grain areas, calculated from the EBSD data presented in Figure 1(b)-(e) were 293 and 1130 μm$^2$ for heat-treatment 1 and 2 respectively. Figure 1(d) and (e) show that the strongest texture components for the recrystallized steel are the {111}||Z (γ-fibre) and the {110}||Y or X directions (α-fibre). Note however, the texture of the sample subjected to heat-treatment 1 is slightly weaker, particularly for the γ-fibre, compared with heat-treatment 2.

To understand the macro-scale response of the samples, reference samples subjected to both heat-treatments were deformed *ex-situ* at a displacement rate of 0.033 mm/s to a strain of approximately 0.05 (total) normal mean strain parallel to the tensile field, as measured by optical Digital Image Correlation (DIC). The results for this are shown in Figure 2.

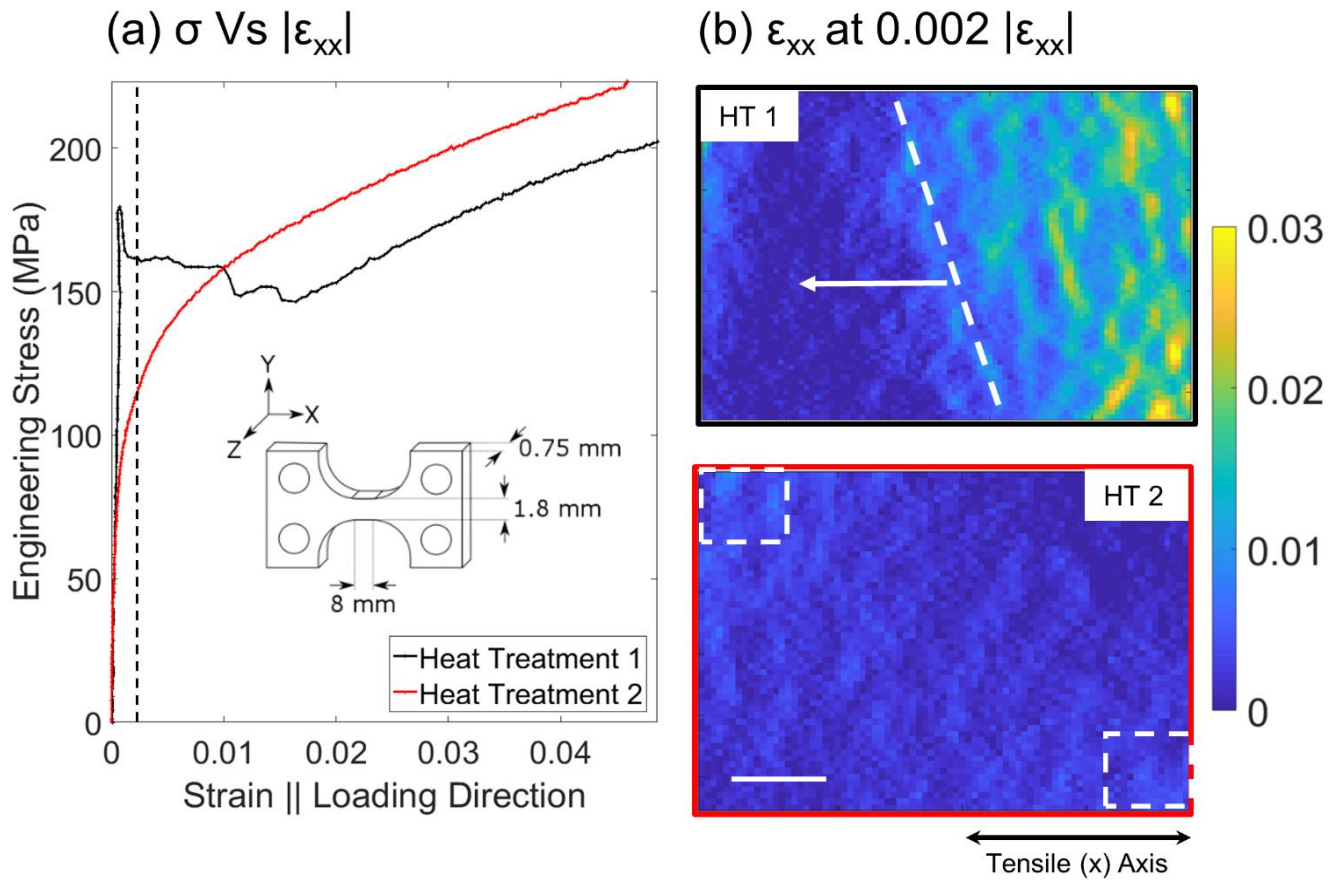

*Figure 2 – Ex-situ engineering stress (σ) vs the mean of the strain field parallel to the x-axis (|εxx|) comparing the macro-scale deformation response of the samples subjected to both heat-treatments with corresponding example strain fields evaluated from DIC at approximately 0.002 mean strain, as shown by the black dashed line on (a). The white scale-bar is 0.5 mm in length. [For interpretation of the colour figures, please see the online version]*

The stress-drops observed in the sample subjected to heat-treatment 1 suggest there is a significant change in the macroscopic response associated with the heat treatments. Load-drops and associated propagation of a Lüders band within the DIC map [47] are shown with the white dashed line and arrow for the sample subjected to heat-treatment 1. Conversely, heat-treatment 2 does not show a clear Lüders band at yield, although there are regions of slightly higher strain at the top left and bottom right corners (shown with white dashed boxes) suggesting that yielding initiated around these regions.

Next, we use the *in-situ* measurements to see if this correlates with a different mechanism at the micrometre-scale. The origins of the different yielding behaviour are mostly likely due to different solute distributions and these are associated with the Ti-based precipitate distributions structures arising from the two heat-treatments. The spatial distribution of the Ti-based precipitates, measured with EDS mapping, along with the raw load-elongation curves from the *in-situ* tensile testing are shown in Figure 3.

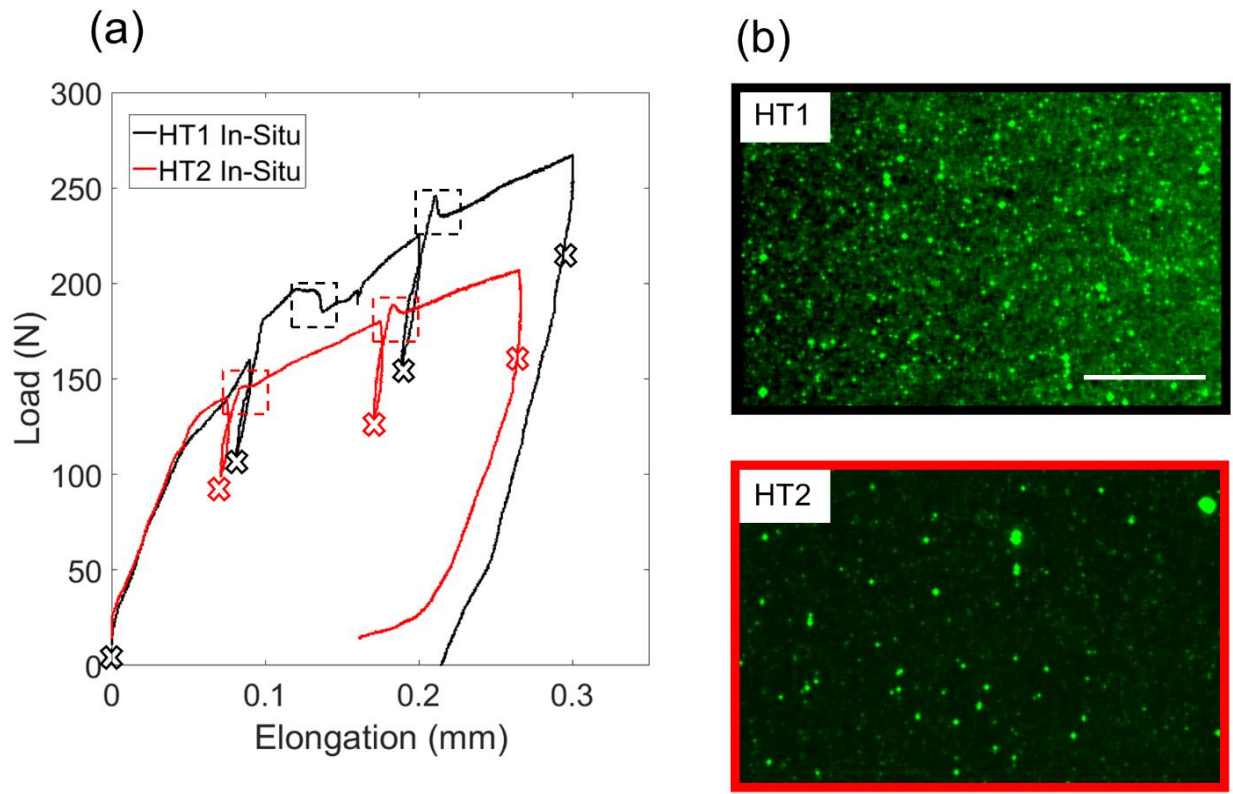

*Figure 3 – (a) Raw load – elongation curves from the in-situ tests showing the effect of interrupting the test for the four HR-EBSD scans (positions marked with a cross) on the reloading yield (highlighted with dashed boxes); (b) qualitative EDS maps showing the effect of each heat-treatment on the resultant Ti-based precipitate distributions. The white scale-bar is 100 μm in length. [For interpretation of the colour figures, please see the online version]*

Firstly, the raw load-elongation curves show a stress-drop after reloading after each HR-EBSD scan in both samples. The magnitude of the stress-drop increases with increasing elongation, i.e. increasing dislocation content for both samples. In these load-hold periods, required for mapping, solute atoms can diffuse and immobilise dislocation structures [48]. Note that the duration of each HR-EBSD scan is approximately 2.5 hours and both *in-situ* tests were conducted at room temperature, implying the solute had sufficient time to diffuse to and pin dislocation structures. Secondly, significant coarsening due to Ostwald-based ripening [35] of the Ti-based precipitates has occurred during the longer soaking time of the sample subjected to heat-treatment 2 (i.e. initially yielded continuously) as shown in Figure 2(b). Therefore, it is likely that solute redistribution occurred: the longer 900°C soak time for the sample that exhibited a continuous yield suggests that more carbon is locked in Ti-based precipitates in the microstructure. As such, the carbon is not dissolved in the matrix and therefore unable to pin mobile dislocations.

The *in-situ* HR-EBSD data is presented in Figure 4 as GND density fields along with a plot of the geometric mean and standard deviation of the GND density with strain.

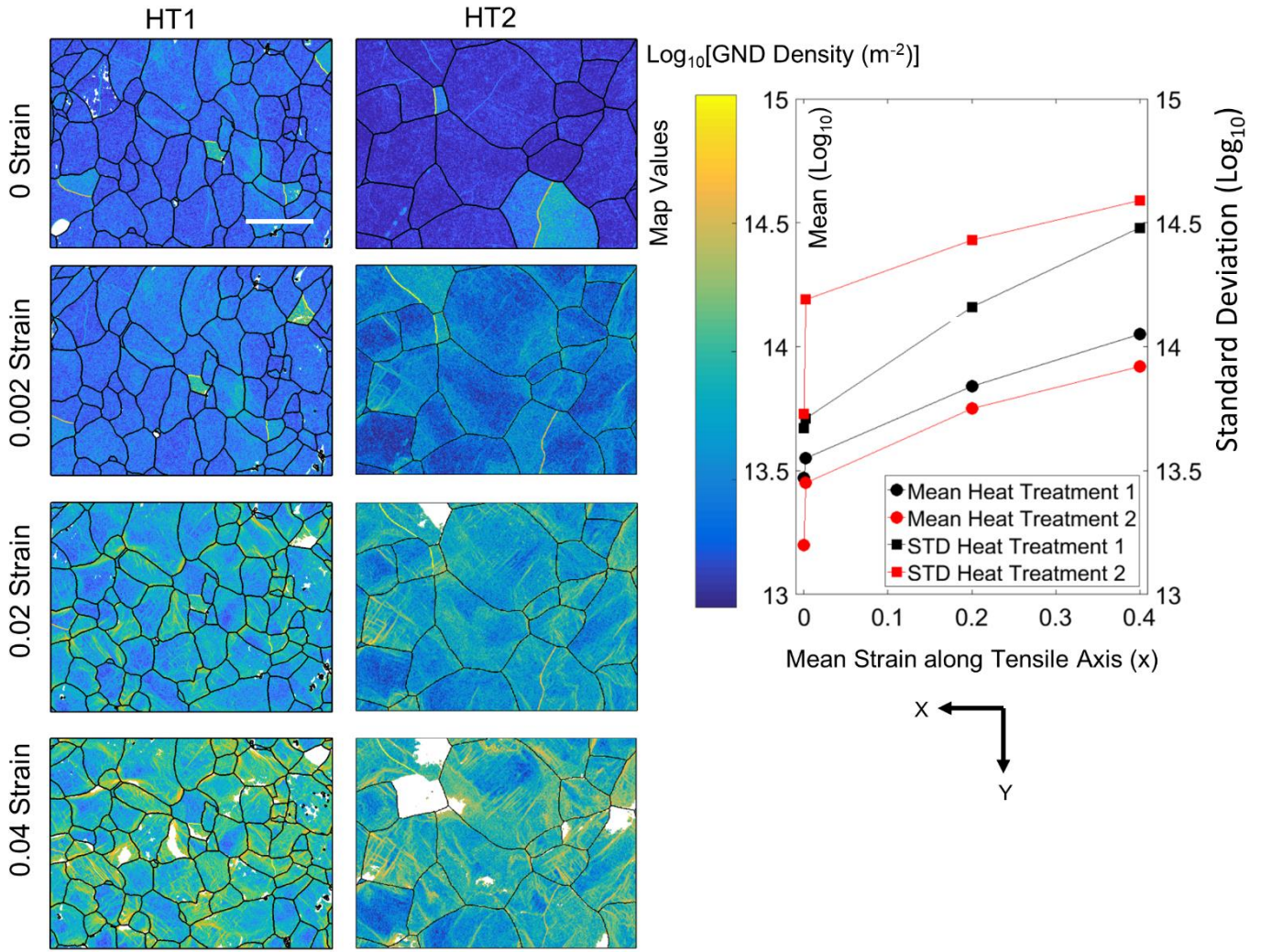

*Figure 4 – Geometrically Necessary Dislocation (GND) density maps as a function of applied mean tensile strain for both samples subjected to heat-treatments 1 & 2 and corresponding plots showing the change in the mean and standard deviation as a function of applied mean tensile strain. The white scale-bar is 50 μm in length. [For interpretation of the colour figures, please see the online version]*

The means and standard deviations of the GND density fields of both samples are seen to increase with strain. The means of the GND density field for heat-treatment 1 are consistently greater for all values of strain compared with heat-treatment 2. Conversely, the standard deviations for heat-treatment 2 are consistently greater for all values of strain compared with heat-treatment 1.

In both samples, the GND density distribution is uniformly low due to recrystallisation. Note however, the GND density field for the first sample subjected to heat-treatment 1 has a slightly higher geometric mean. This is either due to the differences in the recrystallization behaviour between each sample or subtle differences in the final surface finish because of the final ion beam polishing step.

At 0.002 strain (i.e. under macroscopic elastic loading conditions), the sample subjected to heat-treatment 1 retains the flat GND density field observed in the 0 strain case. Conversely, there is significant GND density development for the sample subjected to heat-treatment 2. This correlates well with the observations made with the macro-scale DIC tests for the sample subjected to heat-treatment 2; the strain fields are heterogeneous and develop continuously around the yield-point. However, the observed flat-field of the sample subjected to heat-treatment 1 suggests that the sample was still pre-yield or that the Lüders bands observed in the macro-scale tensile tests had not travelled through the AOI yet.

At 0.02 strain the GND density fields develop significantly for both heat-treatments. There are certain GND density field developments at certain grain boundaries and significant GND density development around triple junctions and at regions with high triple junction density. This is indicative of the neighbourhood effect of the microstructure as noted in [49,50] (i.e. the local locking effect of a triple junction that causes significant GND density development around them). The GND density fields are more sensitive to the effect of local grain neighbourhood as noted by Allain-Bonasso et al. [37] compared with parameters such as grain size.

At 0.04 strain, the GND density fields develop only in magnitude, the distribution within the field shape does not change dramatically. This is indicative of a simple post-yield strain ageing effect observed in macro-scale tensile tests. Note however, the sample subjected to heat-treatment 2 has a significant loss in GND density data in the 0.04 strain case at the top left-hand corner. This is due to local grain rotations that resulted in the grain boundary falling below the grain boundary misorientation threshold of 5º defined in the cross-correlation code and was therefore not defined as a grain boundary. As such, the reference pattern for the cross-correlation changed to one that had a rotation gradient (>~7º) [45] between the test patterns in that area. As such, the GND density data in this area contained a mean angular error and cross-correlation peak-height that were both below the required quality threshold defined [51].

The geometric means as a function of strain are significantly lower for the sample subjected to heat-treatment 2 compared with those of heat-treatment 1 for every value of strain. This is indicative of more homogeneity in the GND density fields for the sample subjected to heat-treatment 2, which is reflected with the continuous yield observed in the *ex-situ* tests. It is likely that this effect due to: (1) the lack of pinning solutes in the ferrite matrix that were removed during heat-treatment 2; (2) the larger mean grain size resulting in a longer mean free path for which a pile-up of dislocations can operate along [20].

One significant advantage of *in-situ* work is to correlate the development of the GND density field within individual grains with the local microstructure. Figure 5 shows example grains selected from the GND density maps presented in Figure 4.

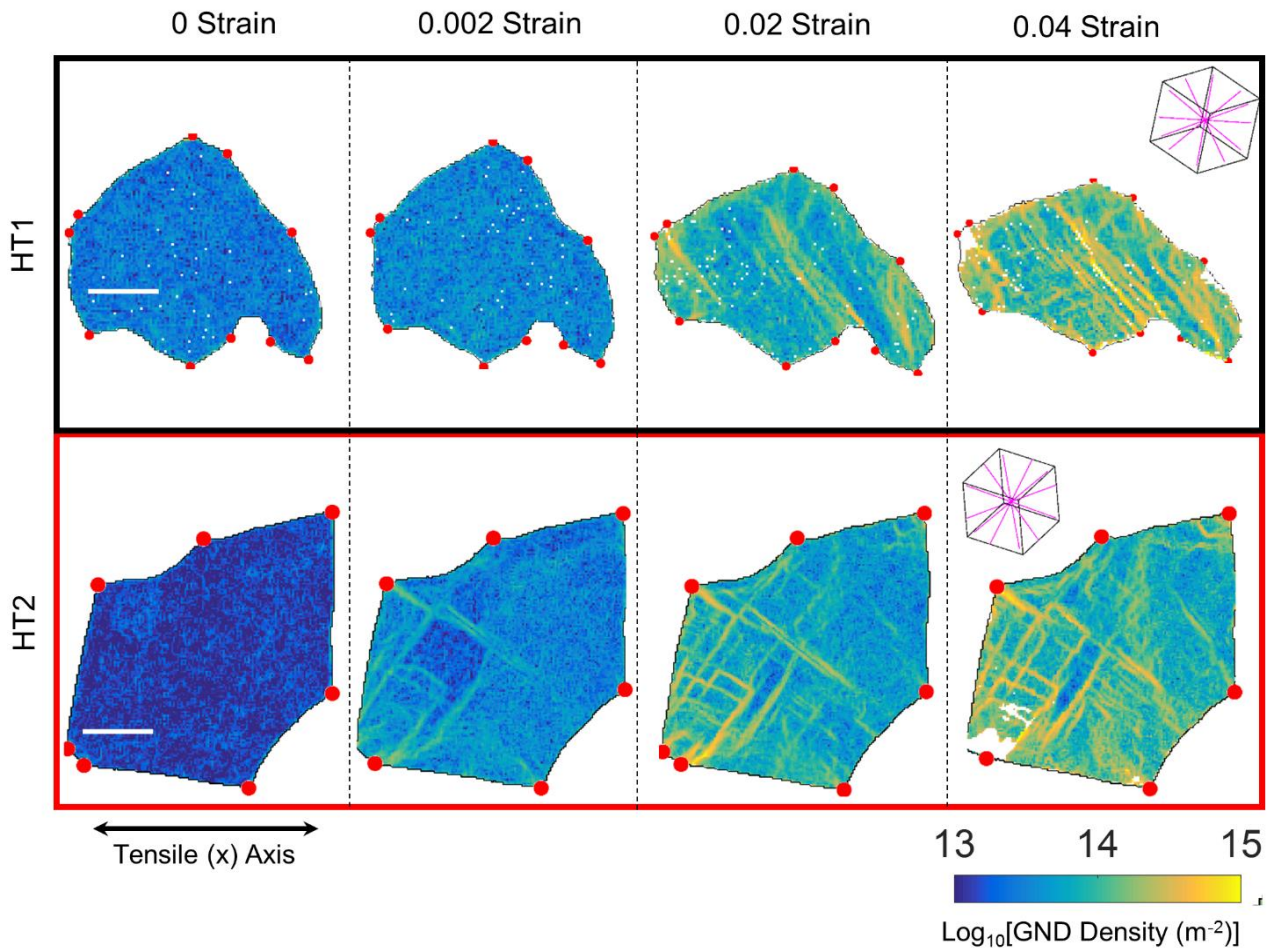

*Figure 5 – Example individual grains showing the development of the GND density fields with strain showing the combined role of triple junctions (marked by red dots) and crystallographic orientation. The mean grain orientation cubes at 0.04 strain with the {110} plane traces plotted in purple is presented. Note the mean grain orientation does not significantly change throughout the entire deformation. The white scale-bars are 10 μm in length. [For interpretation of the colour figures, please see the online version]*

For the grain selected from heat-treatment 1, we see significant heterogeneity develop in the GND density field between 0.002 and 0.02 strain. Specifically, the GND density structures are at their highest magnitude at triple junctions (and along certain grain boundaries). This suggests that compatibility at triple junctions can drive increases in local GND density, and these features extend as linear bands from the triple junctions. This agrees with the findings of Jiang et al. [19] but in our case the linear extent of these fields suggests proximity to the triple junctions should be combined with a direction.

Furthermore, the GND density fields of both grains correlate with the {110} slip planes as shown in the purple plane traces in each orientation cube. However, the GND density fields represent the dislocation content that gives rise to a local strain and rotation gradients, and not necessarily the glissile dislocations that accommodate significant plastic strain. As the grain extracted from the heat-treatment 2 data-set shows, a homogeneous flow hardening is accommodated by the development of an FCC-like cell structure from 0.002 strain onwards.

To summarise:

1. We performed *in-situ* tests combined with HR-EBSD to calculate the GND density field on the same AOI of tensile coupons of two samples of IF steel. The samples had similar grain morphology, texture and from the same rolled sheet. The grain size and perceived concentration of pinning solutes were different.
2. The *in-situ* test was correlated with an *ex-situ* macro-scale Digital Image Correlation. The sample with a perceived higher solute concentration in the ferrite matrix and calculated smaller grain size yielded with stress-drops. Conversely, the other sample exhibited a continuous (FCC-like) yield.
3. Solute redistribution in the microstructure occurred during heat-treatment that resulted in the differences in yield. The sample that initially yielded continuously had coarser Ti-based precipitates compared with the sample that yielded discontinuously due to the longer heat-treatment soaking time. Therefore, any dissolved solute probably diffused to the Ti-based precipitates during heat-treatment resulting in a greater mobile dislocation content to facilitate a continuous yield.

**Acknowledgements:**

J L R Hickey is supported by an ICASE studentship supported by the Engineering and Physical Sciences Research Council and Shell Global Solutions. This work was performed in the Shell Advanced Interfacial Materials Science University Technology Centre. T B Britton thanks the Royal Academy of Engineering for his research fellowship. The authors wish to thank Aran Halder of Tata steel for providing the steel used in this study.

**Data Statement:**

Data will be released to Zenodo upon acceptance of this article & a DOI will be inserted.

**Author Contributions:**

## References

[1]     Statistica, World crude steel production from 2010 to 2017 (in million metric tons), (2017). https://www.statista.com/statistics/267264/world-crude-steel-production/ (accessed December 17, 2018).

[2]     A.H. Cottrell, B.A. Bilby, Dislocation Theory of Yielding and Strain Ageing of Iron, Proc. Phys. Soc. Sect. A. 62 (1949) 49. http://stacks.iop.org/0370-1298/62/i=1/a=308.

[3]     G.P.M. Leyson, L.G. Hector, W.A. Curtin, Solute strengthening from first principles and application to aluminum alloys, Acta Mater. 60 (2012) 3873–3884. doi:10.1016/J.ACTAMAT.2012.03.037.

[4]     H.D. Wang, C. Berdin, M. Mazière, S. Forest, C. Prioul, A. Parrot, P. Le-Delliou, Experimental and numerical study of dynamic strain ageing and its relation to ductile fracture of a C–Mn steel, Mater. Sci. Eng. A. 547 (2012) 19–31. doi:10.1016/J.MSEA.2012.03.069.

[5]     M.R. Wenman, P.R. Chard-Tuckey, Modelling and experimental characterisation of the Lüders strain in complex loaded ferritic steel compact tension specimens, Int. J. Plast. 26 (2010) 1013–1028. doi:10.1016/J.IJPLAS.2009.12.005.

[6]     Y. Estrin, L.P. Kubin, E.C. Aifantis, Introductory remarks to the viewpoint set on propagative plastic instabilities, Scr. Metall. Mater. 29 (1993) 1147–1150. doi:10.1016/0956-716X(93)90100-7.

[7]     R. Sarmah, G. Ananthakrishna, Correlation between band propagation property and the nature of serrations in the Portevin–Le Chatelier effect, Acta Mater. 91 (2015) 192–201. doi:10.1016/J.ACTAMAT.2015.03.027.

[8]     G.P.M. Leyson, W.A. Curtin, L.G. Hector, C.F. Woodward, Quantitative prediction of solute strengthening in aluminium alloys, Nat. Mater. 9 (2010) 750–755. doi:10.1007/978-3-642-37484-5_29.

[9]     D. Blavette, E. Cadel, A. Fraczkiewicz, A. Menand, Three-dimensional atomic-scale imaging of impurity segregation to line defects, Science (80-. ). 286 (1999) 2317–2319.

doi:10.1126/science.286.5448.2317.

[10] M. Hatakeyama, I. Yamagata, Y. Matsukawa, S. Tamura, Direct observation of solute-dislocation interaction on extended edge dislocation in irradiated austenitic stainless steel, Philos. Mag. Lett. 94 (2013) 18–24. doi:10.1080/09500839.2013.853135.

[11] G.P.M. Leyson, W.A. Curtin, L.G. Hector, C.F. Woodward, Quantitative prediction of solute strengthening in aluminium alloys, Nat. Mater. 9 (2010) 750–755. doi:10.1007/978-3-642-37484-5_29.

[12] A.J. Wilkinson, G. Meaden, D.J. Dingley, High-resolution elastic strain measurement from electron backscatter diffraction patterns: new levels of sensitivity., Ultramicroscopy. 106 (2006) 307–13. doi:10.1016/j.ultramic.2005.10.001.

[13] A.J. Wilkinson, T.B. Britton, Strains, planes, and EBSD in materials science, Mater. Today. 15 (2012) 366–376. doi:10.1016/S1369-7021(12)70163-3.

[14] J. Jiang, T. Zhang, F.P.E. Dunne, T. Ben Britton, Deformation compatibility in a single crystalline Ni superalloy, Proc. R. Soc. A Math. Phys. Eng. Sci. 472 (2016) 20150690. doi:10.1098/rspa.2015.0690.

[15] B.E. Jackson, J.J. Christensen, S. Singh, M. De Graef, D.T. Fullwood, E.R. Homer, R.H. Wagoner, Performance of Dynamically Simulated Reference Patterns for Cross-Correlation Electron Backscatter Diffraction, Microsc. Microanal. (2016) 1–14. doi:10.1017/S143192761601148X.

[16] J. Jiang, T. Ben Britton, A.J. Wilkinson, The orientation and strain dependence of dislocation structure evolution in monotonically deformed polycrystalline copper, Int. J. Plast. 69 (2015) 102–117. doi:10.1016/j.ijplas.2015.02.005.

[17] Y. Guo, T.B. Britton, A.J. Wilkinson, Slip band–grain boundary interactions in commercial-purity titanium, Acta Mater. 76 (2014) 1–12. doi:10.1016/j.actamat.2014.05.015.

[18] J. Jiang, T.B. Britton, A.J. Wilkinson, Evolution of dislocation density distributions in copper during tensile deformation, Acta Mater. 61 (2013) 7227–7239. doi:10.1016/j.actamat.2013.08.027.

[19] J. Jiang, T.B. Britton, A.J. Wilkinson, Measurement of geometrically necessary dislocation density

with high resolution electron backscatter diffraction: Effects of detector binning and step size, Ultramicroscopy. 125 (2013) 1–9. doi:10.1016/j.ultramic.2012.11.003.

[20] T.B. Britton, A.J. Wilkinson, Stress fields and geometrically necessary dislocation density distributions near the head of a blocked slip band, Acta Mater. 60 (2012) 5773–5782. doi:10.1016/j.actamat.2012.07.004.

[21] M. Tomita, D. Kosemura, M. Takei, K. Nagata, H. Akamatsu, A. Ogura, Evaluation of strained-silicon by electron backscattering pattern measurement: Comparison study with UV-raman measurement and edge force model calculation, Jpn. J. Appl. Phys. 50 (2011) 010111. doi:10.1143/JJAP.50.010111.

[22] T. Ishido, H. Matsuo, T. Katayama, T. Ueda, K. Inoue, D. Ueda, Depth profiles of strain in AlGaN/GaN heterostructures grown on Si characterized by electron backscatter diffraction technique, IEICE Electron. Express. 4 (2007) 775–781. doi:10.1587/elex.4.775.

[23] M. Tomita, D. Kosemura, M. Takei, K. Nagata, H. Akamatsu, A. Ogura, Evaluation of Strained-Silicon by Electron Backscattering Pattern Measurement: Comparison Study with UV-Raman Measurement and Edge Force Model Calculation, Jpn. J. Appl. Phys. 50 (2011) 010111. doi:10.1143/JJAP.50.010111.

[24] M.D. Vaudin, W.A. Osborn, L.H. Friedman, J.M. Gorham, V. Vartanian, R.F. Cook, Designing a standard for strain mapping: HR-EBSD analysis of SiGe thin film structures on Si, Ultramicroscopy. 148 (2015) 94–104. doi:10.1016/J.ULTRAMIC.2014.09.007.

[25] D. Wallis, L.N. Hansen, T. Ben Britton, A.J. Wilkinson, Dislocation Interactions in Olivine Revealed by HR-EBSD, J. Geophys. Res. Solid Earth. 122 (2017) 7659–7678. doi:10.1002/2017JB014513.

[26] D. Wallis, L.N. Hansen, T. Ben Britton, A.J. Wilkinson, Geometrically necessary dislocation densities in olivine obtained using high-angular resolution electron backscatter diffraction, Ultramicroscopy. 168 (2016) 34–45. doi:10.1016/J.ULTRAMIC.2016.06.002.

[27] T.B. Britton, J. Jiang, P.S. Karamched, A.J. Wilkinson, Probing Deformation and Revealing Microstructural Mechanisms with Cross-Correlation-Based, High-Resolution Electron Backscatter Diffraction, Jom. 65 (2013) 1245–1253. doi:10.1007/s11837-013-0680-6.

[28] P.D. Littlewood, T.B. Britton, A.J. Wilkinson, Geometrically necessary dislocation density distributions in Ti–6Al–4V deformed in tension, Acta Mater. 59 (2011) 6489–6500. doi:10.1016/j.actamat.2011.07.016.

[29] A.J. Wilkinson, D. Randman, Determination of elastic strain fields and geometrically necessary dislocation distributions near nanoindents using electron backscatter diffraction, Philos. Mag. 90 (2010) 1159–1177. doi:10.1080/14786430903304145.

[30] T.J. Ruggles, D.T. Fullwood, Estimations of bulk geometrically necessary dislocation density using high resolution EBSD, Ultramicroscopy. 133 (2013) 8–15. doi:10.1016/j.ultramic.2013.04.011.

[31] W. Pantleon, Resolving the geometrically necessary dislocation content by conventional electron backscattering diffraction, Scr. Mater. 58 (2008) 994–997. doi:10.1016/j.scriptamat.2008.01.050.

[32] M.R. Staker, D.L. Holt, The Dislocation Cell Size and Dislocation Density in Copper Deformed at Temperatures between 25 and 700 celcius, Acta Metall. 20 (1972) 569–579.

[33] M.J. McLean, W.A. Osborn, In-situ elastic strain mapping during micromechanical testing using EBSD, Ultramicroscopy. 185 (2018) 21–26. doi:10.1016/J.ULTRAMIC.2017.11.007.

[34] J. Ast, G. Mohanty, Y. Guo, J. Michler, X. Maeder, In situ micromechanical testing of tungsten micro-cantilevers using HR-EBSD for the assessment of deformation evolution, Mater. Des. 117 (2017) 265–266. doi:10.1016/J.MATDES.2016.12.052.

[35] S. Hoile, Processing and properties of mild interstitial free steels, Mater. Sci. Technol. 16 (2000) 1079–1093. doi:10.1179/026708300101506902.

[36] B.L. Li, A. Godfrey, Q.C. Meng, Q. Liu, L.N. Hansen, Microstructural evolution of IF-steel during cold rolling, Acta Mater. 52 (2004) 1069–1081. doi:10.1016/j.actamat.2003.10.040.

[37] N. Allain-Bonasso, F. Wagner, S. Berbenni, D.P. Field, A study of the heterogeneity of plastic deformation in IF steel by EBSD, Mater. Sci. Eng. A. 548 (2012) 56–63. doi:10.1016/j.msea.2012.03.068.

[38] B.J. Duggan, Y.Y. Tse, G. Lam, M.Z. Quadir, Deformation and Recrystallization of Interstitial Free (IF) Steel, Mater. Manuf. Process. 26 (2011) 51–57. doi:10.1080/10426910903202237.

[39]  S. Gao, M. Chen, M. Joshi, A. Shibata, N. Tsuji, Yielding behavior and its effect on uniform elongation in if steel with various grain sizes, J. Mater. Sci. 49 (2014) 6536–6542. doi:10.1007/s10853-014-8233-0.

[40]  S. Gao, A. Shibata, M. Chen, N. Park, N. Tsuji, Correlation between Continuous/Discontinuous Yielding and Hall-Petch Slope in High Purity Iron, Mater. Trans. 55 (2014) 69–72. doi:10.2320/matertrans.MA201326.

[41]  J.S. and X. Wang, Comparison of Precipitate Behaviors in Ultra-Low Carbon, Titanium-Stabilized Interstitial Free Steel Sheets under Different Annealing Processes, J. Mater. Eng. Perform. 8 (1999) 641–648. doi:10.1361/105994999770346396.

[42]  J. Jiang, J. Yang, T. Zhang, F.P.E. Dunne, T.B. Britton, On the mechanistic basis of fatigue crack nucleation in Ni superalloy containing inclusions using high resolution electron backscatter diffraction, Acta Mater. 97 (2015) 367–379. doi:10.1016/j.actamat.2015.06.035.

[43]  J. Witcomb, Dislocation Cell Structure Relation, 299 (1974) 299–304.

[44]  F. Bachmann, R. Hielscher, H. Schaeben, Texture Analysis with MTEX – Free and Open Source Software Toolbox, Solid State Phenom. 160 (2010) 63–68. doi:10.4028/www.scientific.net/SSP.160.63.

[45]  T.B. Britton, A.J. Wilkinson, High resolution electron backscatter diffraction measurements of elastic strain variations in the presence of larger lattice rotations., Ultramicroscopy. 114 (2012) 82–95. doi:10.1016/j.ultramic.2012.01.004.

[46]  J.F. Nye, Some geometrical relations in dislocated crystals, Acta Metall. 1 (1953) 153–162. doi:10.1016/0001-6160(53)90054-6.

[47]  W. Leslie, The Physical Metallurgy of Steels, 1st ed., New York, 1981.

[48]  R.B. McLellan, M.L. Wasz, Carbon diffusivity in B.C.C. iron, J. Phys. Chem. Solids. 54 (1993) 583–586. doi:10.1016/0022-3697(93)90236-K.

[49]  G. Winther, J.P. Wright, S. Schmidt, J. Oddershede, Grain interaction mechanisms leading to intragranular orientation spread in tensile deformed bulk grains of interstitial-free steel, Int. J. Plast. 88 (2017) 108–125. doi:10.1016/j.ijplas.2016.10.004.

[50] H. Abdolvand, J. Wright, A.J. Wilkinson, Strong grain neighbour effects in polycrystals, Nat. Commun. 9 (2018) 171. doi:10.1038/s41467-017-02213-9.

[51] T.B. Britton, A.J. Wilkinson, Measurement of residual elastic strain and lattice rotations with high resolution electron backscatter diffraction., Ultramicroscopy. 111 (2011) 1395–404. doi:10.1016/j.ultramic.2011.05.007.